\documentclass{article}
\usepackage{amsfonts}

\usepackage{amsmath}

\newtheorem{theorem}{Theorem}

\newtheorem{proposition}[theorem]{Proposition}
\newtheorem{remark}[theorem]{Remark}

\newenvironment{proof}[1][Proof]{\noindent\textbf{#1.} }{\ \rule{0.5em}{0.5em}}
\input{tcilatex}

\begin{document}

\title{Towards a Direct Method for the Analyticity of the Pressure for
Certain Classical Unbounded Spin Systems }
\author{Assane Lo \\
The University of Arizona}
\maketitle

\begin{abstract}
The aim of this paper is to study direct methods for the analyticity of the
pressure for certain classical unbounded spin models. We provide a
representation in terms of the Witten Laplacian on one-forms of the
nth-derivative of the pressure as function of some order parameter t. The
technique involves the formula for the covariance introduced by B. Helffer
and J. Sjostrand.
\end{abstract}

\section{Introduction}

As already mentioned in [66], The methods for investigating critical
phenomena for certain physical systems took an interesting direction when
powerful and sophisticated PDE techniques were introduced. The methods are
generally based on the analysis of suitable differential operators%
\begin{equation*}
\mathbf{W}_{\Phi }^{\left( 0\right) }=\left( \mathbf{-\Delta +}\frac{\left| 
\mathbf{\nabla }\Phi \right| ^{2}}{4}-\frac{\mathbf{\Delta }\Phi }{2}\right) 
\end{equation*}%
and%
\begin{equation*}
\mathbf{W}_{\Phi }^{\left( 1\right) }=\mathbf{-\Delta +}\frac{\left| \mathbf{%
\nabla }\Phi \right| ^{2}}{4}-\frac{\mathbf{\Delta }\Phi }{2}+\mathbf{Hess}%
\Phi .
\end{equation*}%
These are in some sense deformations of the standard Laplace Beltrami
operator. They are commonly called Witten Laplacians, and were first
introduced by Edward Witten, [18] in 1982 in the context of Morse theory for
the study of topological invariants of compact Riemannian manifolds. In
1994, Bernard Helffer and Johannes Sj\"{o}strand [8] introduced two elliptic
differential operators%
\begin{equation*}
A_{\Phi }^{(0)}:=-\mathbf{\Delta }+\mathbf{\nabla }\Phi \cdot \mathbf{\nabla 
}
\end{equation*}%
and 
\begin{equation*}
A_{\Phi }^{(1)}:=-\mathbf{\Delta }+\mathbf{\nabla }\Phi \cdot \mathbf{\nabla
+Hess}\Phi 
\end{equation*}%
sometimes called Helffer-Sj\"{o}strand operators serving to get direct
method for the study of integrals and operators in high dimensions of the
type that appear in Statistical Mechanics and Euclidean Field Theory. In
1996, Johannes Sj\"{o}strand [13] observed that these so-called Helffer-Sj%
\"{o}strand operators were in fact equivalent to Witten's Laplacians. Since
then, there have been significant advances in the use of these Laplacians to
study the thermodynamic behavior of quantities related to the Gibbs measure $%
Z^{-1}e^{-\Phi }dx.$

Numerous techniques have been developed in the study of integrals associated
with the equilibrium Gibbs state for certain unbounded spins systems. One of
the most striking results is an exact formula for the covariance of two
functions in terms of the Witten Laplacian on one forms leading to
sophisticated methods for estimating the correlation functions of a random
field. As mentioned in [10], this formula is in some sense a stronger and
more flexible version of the Brascamp-Lieb inequality [1]. The formula may
be written as follow:%
\begin{equation}
\mathbf{cov}(f,g)=\int \left( A_{\Phi }^{(1)^{-1}}\mathbf{\nabla }f\cdot 
\mathbf{\nabla }g\right) e^{-\Phi (x)}dx.
\end{equation}

We attempt in these notes, to study a direct method for the analyticity of
the pressure for certain classical convex unbounded spin systems. It is
central in Statistical Mechanics to study the differentiability or even the
analyticity of the pressure with respect to some distinguished thermodynamic
parameters such as temperature, chemical potential or external field. In
fact the analytic behavior of the pressure is the classical thermodynamic
indicator for the absence or existence of phase transition. The most famous
result on the analyticity of the pressure is the circle theorem of Lee and
Yang [28]. This theorem asserts the following: consider a $\left\{
-1,1\right\} -$valued spin system with ferromagnetic pair interaction and
external field $h$ and regard the quantity $z=e^{h}$ as a complex parameter,
then all zeroes of all partition functions (with free boundary condition),
considered as functions of $z$ lie in the complex unit circle. This theorem
readily implies that the pressure is an analytic function of $h$ in the
region $h>0$ and $h<0.$ Heilmann [29] showed that the assumption of pair
interaction is necessary. A transparent approach to the circle theorem was
found by Asano [30] and developed further by Ruelle [31],[32], Slawny [33],
and Gruber et al [34]. Griffiths [35] and Griffiths-Simon [36] found a
method of extending the Lee-Yang theorem to real-valued spin systems with a
particular type of a priory measure. Newman [37] proved the Lee-Yang theorem
for every a priory measure which satisfies this theorem in the particular
case of no interaction. Dunlop [38],[39] studied the zeroes of the partition
functions for the plane rotor model. A general Lee-Yang theorem for
multicomponent systems was finally proved by Lieb and Sokal [40]. For
further references see Glimm and Jaffe [41].\newline
The Lee-Yang theorem and its variants depend on the ferromagnetic character
of the interaction. There are various other way of proving the infinite
differentiability or the analyticity of the pressure for (ferromagnetic and
non ferromagnetic) systems at high temperatures, or at low temperatures, or
at large external fields. Most of these take advantage of a sufficiently
rapid decay of correlations and /or cluster expansion methods. Here is a
small sample of relevant references. Bricmont, Lebowitz and Pfister [42],
Dobroshin [43], Dobroshin and Sholsman [44],[45], Duneau et al
[46],[47],[48], Glimm and Jaffe [41],[49], Israel [50], Kotecky and Preiss
[51], Kunz [52], Lebowitz [53],[54], Malyshev [55], Malychev and Milnos [56]
and Prakash [57]. M. Kac and J.M. Luttinger [58] obtained a formula for the
pressure in terms of irreducible distribution functions.

In this present study, we propose a new way of analyzing the analyticity of
the pressure for certain unbounded models through a representation by means
of the Witten Laplacians of the remainder of the Taylor series expansion.
The methods known up to now rely on complicated indirect arguments.

\section{Towards the analyticity of the Pressure}

Let $\Lambda $\ be a finite domain in $\mathbb{Z}^{d}\;(d\geq 1)$ and
consider the Hamiltonian of the phase space given by,%
\begin{equation}
\Phi (x)=\Phi _{\Lambda }(x)=\frac{x^{2}}{2}+\Psi (x),\;\ \ \ \ \;x\in 
\mathbb{R}^{\Lambda }.
\end{equation}%
where%
\begin{equation}
\left| \partial ^{\alpha }\mathbf{\nabla }\Psi \right| \leq C_{\alpha },\;\
\ \ \ \ \forall \alpha \in \mathbb{N}^{\left| \Lambda \right| },
\end{equation}%
\begin{equation}
\mathbf{Hess}\Phi (x)\geq \delta _{o},\;\;\;\;\;\;\;\;\;\;0<\delta _{o}<1.
\end{equation}%
Let $g$ is a smooth function on $\mathbb{R}^{\Gamma }$ with lattice support $%
S_{g}=\Gamma .$ We identified with $\tilde{g}$ defined on $\mathbb{R}%
^{\Lambda }$ by 
\begin{equation}
\tilde{g}(x)=g(x_{\Gamma })\text{ \ \ where }x=\left( x_{i}\right) _{i\in
\Lambda }\text{ \ and }x_{\Gamma }=\left( x_{i}\right) _{i\in \Gamma }
\end{equation}%
and satisfying%
\begin{equation}
\left| \partial ^{\alpha }\mathbf{\nabla }g\right| \leq C_{\alpha
}\;\;\;\;\;\;\;\;\forall \alpha \in \mathbb{N}^{\left| \Gamma \right| }
\end{equation}%
Under the additional assumptions that $\Psi $ is compactly supported in $%
\mathbb{R}^{\Lambda }$ and $g$ is compactly supported in $\mathbb{R}^{\Gamma
},$ it was proved in [66] (see also [8]) that the equation 
\begin{equation*}
\left\{ 
\begin{tabular}{l}
$-\mathbf{\Delta }f+\mathbf{\nabla }\Phi \cdot \mathbf{\nabla }%
f=g-\left\langle g\right\rangle $ \\ 
$\left\langle f\right\rangle _{L^{2}(\mu )}=0$%
\end{tabular}%
\right. 
\end{equation*}%
has a unique smooth solution satisfying $\mathbf{\nabla }^{k}f(x)\rightarrow
0$ as $\left| x\right| \rightarrow \infty $ for every $k\geq 1$.

Recall also that $\mathbf{\nabla }f$ is a solution of the system%
\begin{equation}
\left( -\mathbf{\Delta +\nabla }\Phi \cdot \mathbf{\nabla }\right) \mathbf{%
\nabla }f+\mathbf{Hess}\Phi \mathbf{\nabla }f=\mathbf{\nabla }g\;\;\;\;\text{%
in }\;\mathbb{R}^{\Lambda }.
\end{equation}%
As in [66] and [8], these assumptions will be relaxed later on.

Let 
\begin{equation}
\Phi _{\Lambda }^{t}(x)=\Phi (x)-tg(x),
\end{equation}%
where $x=(x_{i})_{i\in \Lambda }$, and assume additionally that $g$
satisfies 
\begin{equation}
\mathbf{Hess}g\leq C.
\end{equation}%
We consider the following perturbation \ 
\begin{equation}
\theta _{\Lambda }(t)=\log \left[ \int dxe^{-\Phi _{\Lambda }^{t}(x)}\right]
.
\end{equation}%
Denote by 
\begin{equation}
Z_{t}=\int dxe^{-\Phi _{\Lambda }^{t}(x)}
\end{equation}%
and 
\begin{equation}
<\cdot >_{t,\Lambda }=\frac{\int \cdot \;dxe^{-\Phi _{\Lambda }^{t}(x)}}{%
Z_{t}}.
\end{equation}

\section{Parameter Dependency of the Solution}

From the assumptions made on $\Phi $ and $g,$ it is easy to see that there
exists $T>0$ such that or every $t\in \lbrack 0,T)$, $\Phi _{\Lambda
}^{t}(x) $ satisfies all the assumptions required for the solvability,
regularity and asymptotic behavior of the solution $f(t)$ associated with
the potential $\Phi _{\Lambda }^{t}(x).$ Thus, each $t\in \lbrack 0,T)$ is
associated with a unique $C^{\infty }-$solution, $f(t)$ of the equation 
\begin{equation*}
\left\{ 
\begin{tabular}{l}
$A_{\Phi _{\Lambda }^{t}}^{(0)}f(t)=g-\left\langle g\right\rangle
_{_{L^{2}(\mu )}}$ \\ 
$\left\langle f(t)\right\rangle _{L^{2}(\mu )}=0.$%
\end{tabular}%
\right.
\end{equation*}%
Hence,%
\begin{equation}
A_{\Phi _{\Lambda }^{t}}^{(1)}\mathbf{v}(t)=\mathbf{\nabla }g
\end{equation}%
where $\mathbf{v}(t)=\mathbf{\nabla }f(t).$ Notice that the map 
\begin{equation*}
t\longmapsto \mathbf{v}(t)
\end{equation*}%
is well defined and 
\begin{equation*}
\left\{ \mathbf{v}(t):t\in \lbrack 0,T)\right\}
\end{equation*}%
is a family of smooth solutions on $\mathbb{R}^{\Lambda }$ satisfying 
\begin{equation*}
\partial ^{\alpha }\mathbf{v}(t)\rightarrow 0\;\;\;\text{as }\left| x\right|
\rightarrow \infty \ \ \ \ \ \ \ \forall \alpha \in \mathbb{N}^{\left|
\Lambda \right| }\ \text{\ and for each }t\in \lbrack 0,T)
\end{equation*}%
and corresponding to the family of potential 
\begin{equation}
\left\{ \Phi _{\Lambda }^{t}:t\in \lbrack 0,T)\right\} .
\end{equation}%
Let us now verify that $\mathbf{v}$ is a smooth function of $t\in (0,T).$We
need to prove that for each $t\in (0,T),$ the limit%
\begin{equation*}
\lim\limits_{\varepsilon \rightarrow 0}\frac{\mathbf{v}(t+\varepsilon )-%
\mathbf{v}(t)}{\varepsilon }
\end{equation*}%
exists. Let 
\begin{equation*}
\mathbf{v}^{\varepsilon }(t)=\frac{\mathbf{v}(t+\varepsilon )-\mathbf{v}(t)}{%
\varepsilon }.
\end{equation*}%
We use a technique based on regularity estimates to get a uniform control of 
$\mathbf{v}^{\varepsilon }(t)$ with respect to $\varepsilon .$

With $\varepsilon $ small enough, we have%
\begin{eqnarray*}
0 &=&-\mathbf{\Delta }\left[ \dfrac{\mathbf{v}(t+\varepsilon )-\mathbf{v}(t)%
}{\varepsilon }\right] +\dfrac{\mathbf{\nabla }\Phi ^{t+\varepsilon }\cdot 
\mathbf{\nabla v}(t+\varepsilon )-\mathbf{\nabla }\Phi ^{t}\cdot \mathbf{%
\nabla v}(t)}{\varepsilon } \\
&&+\dfrac{\mathbf{Hess}\Phi ^{t+\varepsilon }\mathbf{v}(t+\varepsilon )-%
\mathbf{Hess}\Phi ^{t}\mathbf{v}(t)}{\varepsilon }.
\end{eqnarray*}%
Equivalently,%
\begin{eqnarray*}
&&-\mathbf{\Delta }\left[ \dfrac{\mathbf{v}(t+\varepsilon )-\mathbf{v}(t)}{%
\varepsilon }\right] +\dfrac{\mathbf{\nabla }\Phi ^{t+\varepsilon }\cdot 
\mathbf{\nabla }\left[ \mathbf{v}(t+\varepsilon )-\mathbf{v}(t)\right] }{%
\varepsilon } \\
&&+\mathbf{Hess}\Phi ^{t+\varepsilon }\left( \dfrac{\mathbf{v}(t+\varepsilon
)-\mathbf{v}(t)}{\varepsilon }\right) \\
&=&-\left( \dfrac{\mathbf{Hess}\Phi ^{t+\varepsilon }-\mathbf{Hess}\Phi ^{t}%
}{\varepsilon }\right) \mathbf{v}(t)-\left( \dfrac{\mathbf{\nabla }\Phi
^{t+\varepsilon }-\mathbf{\nabla }\Phi ^{t}}{\varepsilon }\right) \cdot 
\mathbf{\nabla v}(t)
\end{eqnarray*}%
and%
\begin{eqnarray*}
&&-\mathbf{\Delta v}^{\varepsilon }(t)+\mathbf{\nabla }\Phi ^{t+\varepsilon
}\cdot \mathbf{\nabla v}^{\varepsilon }(t)+\mathbf{Hess}\Phi ^{t+\varepsilon
}\mathbf{v}^{\varepsilon }(t) \\
&=&-\left( \dfrac{\mathbf{Hess}\Phi ^{t+\varepsilon }-\mathbf{Hess}\Phi ^{t}%
}{\varepsilon }\right) \mathbf{v}(t)-\left( \dfrac{\mathbf{\nabla }\Phi
^{t+\varepsilon }-\mathbf{\nabla }\Phi ^{t}}{\varepsilon }\right) \cdot 
\mathbf{\nabla v}(t).
\end{eqnarray*}%
Let $\mathbf{w}(t)$ be the unique $C^{\infty }-$solution of the system 
\begin{equation}
-\mathbf{\Delta w}(t)+\mathbf{\nabla }\Phi ^{t}\cdot \mathbf{\nabla w}(t)+%
\mathbf{Hess}\Phi ^{t}\mathbf{w}(t)=\mathbf{Hess}g\mathbf{v}(t)-\mathbf{%
\nabla }g\cdot \mathbf{\nabla v}(t).
\end{equation}%
Combining the last two systems above, we get\newline
\begin{equation}
\begin{array}{c}
-\mathbf{\Delta }\left[ \mathbf{w}(t)-\mathbf{v}^{\varepsilon }(t)\right] +%
\mathbf{\nabla }\Phi ^{t}\cdot \mathbf{\nabla }\left[ \mathbf{w}(t)-\mathbf{v%
}^{\varepsilon }(t)\right] +\mathbf{Hess}\Phi ^{t}\left[ \mathbf{w}(t)-%
\mathbf{v}^{\varepsilon }(t)\right] \\ 
=\mathbf{Hess}g\mathbf{v}(t)-\mathbf{\nabla }g\cdot \mathbf{\nabla v}%
(t)+\left( \dfrac{\mathbf{Hess}\Phi ^{t+\varepsilon }-\mathbf{Hess}\Phi ^{t}%
}{\varepsilon }\right) \mathbf{v}(t) \\ 
+\left( \dfrac{\mathbf{\nabla }\Phi ^{t+\varepsilon }-\mathbf{\nabla }\Phi
^{t}}{\varepsilon }\right) \cdot \mathbf{\nabla v}(t)+\left( \mathbf{\nabla }%
\Phi ^{t+\varepsilon }-\mathbf{\nabla }\Phi ^{t}\right) \cdot \mathbf{\nabla
v}^{\varepsilon }(t) \\ 
+\left( \mathbf{Hess}\Phi ^{t+\varepsilon }-\mathbf{Hess}\Phi ^{t}\right) 
\mathbf{v}^{\varepsilon }(t).%
\end{array}%
\end{equation}%
Now using the unitary transformation $U_{\Phi ^{t}},$ we get%
\begin{equation}
\begin{array}{c}
\left( \mathbf{-\Delta +}\dfrac{\left| \mathbf{\nabla }\Phi ^{t}\right| ^{2}%
}{4}-\dfrac{\mathbf{\Delta }\Phi ^{t}}{2}\right) \left( \mathbf{w}(t)-%
\mathbf{v}^{\varepsilon }(t)\right) e^{-\Phi ^{t}/2} \\ 
+\mathbf{Hess}\Phi ^{t}\left( \mathbf{w}(t)-\mathbf{v}^{\varepsilon
}(t)\right) e^{-\Phi ^{t}/2} \\ 
=o_{\varepsilon }(1)e^{-\Phi ^{t}/2}+[\left( \mathbf{\nabla }\Phi
^{t+\varepsilon }-\mathbf{\nabla }\Phi ^{t}\right) \cdot \mathbf{\nabla v}%
^{\varepsilon }(t) \\ 
+\left( \mathbf{Hess}\Phi ^{t+\varepsilon }-\mathbf{Hess}\Phi ^{t}\right) 
\mathbf{v}^{\varepsilon }(t)]e^{-\Phi ^{t}/2}%
\end{array}%
\end{equation}%
Next, we propose to estimate the last two terms of the right hand side of
this equation.

Again using the unitary transformation $U_{\Phi ^{t+\varepsilon },}$ we
reduce the system\newline
\begin{equation}
\begin{array}{c}
-\mathbf{\Delta v}^{\varepsilon }(t)+\mathbf{\nabla }\Phi ^{t+\varepsilon
}\cdot \mathbf{\nabla v}^{\varepsilon }(t)+\mathbf{Hess}\Phi ^{t+\varepsilon
}\mathbf{v}^{\varepsilon }(t) \\ 
=-\left( \dfrac{\mathbf{Hess}\Phi ^{t+\varepsilon }-\mathbf{Hess}\Phi ^{t}}{%
\varepsilon }\right) \mathbf{v}(t) \\ 
-\left( \dfrac{\mathbf{\nabla }\Phi ^{t+\varepsilon }-\mathbf{\nabla }\Phi
^{t}}{\varepsilon }\right) \cdot \mathbf{\nabla v}(t)%
\end{array}%
\end{equation}%
into%
\begin{equation}
\left. 
\begin{array}{c}
\left( \mathbf{-\Delta +}\dfrac{\left| \mathbf{\nabla }\Phi ^{t+\varepsilon
}\right| ^{2}}{4}-\dfrac{\mathbf{\Delta }\Phi ^{t+\varepsilon }}{2}\right) 
\mathbf{V}^{\varepsilon }+\mathbf{Hess}\Phi ^{t+\varepsilon }\mathbf{V}%
^{\varepsilon }= \\ 
-\left( \dfrac{\mathbf{Hess}\Phi ^{t+\varepsilon }-\mathbf{Hess}\Phi ^{t}}{%
\varepsilon }\right) \mathbf{v}(t)e^{-\Phi ^{t+\varepsilon }/2} \\ 
-\left( \dfrac{\mathbf{\nabla }\Phi ^{t+\varepsilon }-\mathbf{\nabla }\Phi
^{t}}{\varepsilon }\right) \cdot \mathbf{\nabla v}(t)e^{-\Phi
^{t+\varepsilon }/2}%
\end{array}%
\right.
\end{equation}%
where $\mathbf{V}^{\varepsilon }=\mathbf{v}^{\varepsilon }(t)e^{-\Phi
^{t+\varepsilon }/2}.$ Taking scalar product with $\mathbf{V}^{\varepsilon }$
on both sides of this last equality and integrating, we get\newline
\begin{equation}
\left. 
\begin{array}{c}
\left\| \left( \partial _{x}\mathbf{+}\dfrac{\mathbf{\nabla }\Phi
^{t+\varepsilon }}{2}\right) \mathbf{V}^{\varepsilon }\right\|
_{L^{2}}^{2}+\int \mathbf{Hess}\Phi ^{t+\varepsilon }\mathbf{V}^{\varepsilon
}\cdot \mathbf{V}^{\varepsilon }dx= \\ 
-\int \left[ \left( \dfrac{\mathbf{Hess}\Phi ^{t+\varepsilon }-\mathbf{Hess}%
\Phi ^{t}}{\varepsilon }\right) \mathbf{v}(t)e^{-\Phi ^{t+\varepsilon }/2}%
\right] \cdot \mathbf{V}^{\varepsilon }dx \\ 
-\left[ \int \left( \dfrac{\mathbf{\nabla }\Phi ^{t+\varepsilon }-\mathbf{%
\nabla }\Phi ^{t}}{\varepsilon }\right) \cdot \mathbf{\nabla v}(t)e^{-\Phi
^{t+\varepsilon }/2}\right] \cdot \mathbf{V}^{\varepsilon }dx%
\end{array}%
\right.
\end{equation}%
Now using the uniform strict convexity on the left hand side and
Cauchy-Schwartz on the right hand side, we obtain 
\begin{equation}
\left\| \mathbf{V}^{\varepsilon }\right\| _{B^{0}}\leq C\text{ \ \ \ \ for
small enough }\varepsilon .
\end{equation}%
We then deduce that 
\begin{equation}
\left( \mathbf{-\Delta +}\frac{\left| \mathbf{\nabla }\Phi ^{t+\varepsilon
}\right| ^{2}}{4}\right) \mathbf{V}^{\varepsilon }=\tilde{q}_{\varepsilon }
\end{equation}%
where%
\begin{equation}
\left. 
\begin{array}{c}
\tilde{q}_{\varepsilon }=-\left( \dfrac{\mathbf{Hess}\Phi ^{t+\varepsilon }-%
\mathbf{Hess}\Phi ^{t}}{\varepsilon }\right) \mathbf{v}(t)e^{-\Phi
^{t+\varepsilon }/2}-\left( \dfrac{\mathbf{\nabla }\Phi ^{t+\varepsilon }-%
\mathbf{\nabla }\Phi ^{t}}{\varepsilon }\right) \cdot \mathbf{\nabla v}%
(t)e^{-\Phi ^{t+\varepsilon }/2} \\ 
\mathbf{+}\dfrac{\mathbf{\Delta }\Phi ^{t+\varepsilon }}{2}\mathbf{V}%
^{\varepsilon }-\mathbf{Hess}\Phi ^{t+\varepsilon }\mathbf{V}^{\varepsilon }%
\end{array}%
\right.
\end{equation}%
is bounded in $B^{0}$ uniformly with respect to $\varepsilon $ for $%
\varepsilon $ small enough.\newline
Now taking scalar product with $\mathbf{V}^{\varepsilon }$ on both sides of $%
\left( 28\right) $ and integrating by parts, we obtain%
\begin{equation}
\left\| \mathbf{\nabla V}^{\varepsilon }\right\| _{L^{2}}^{2}+\left\| \frac{%
\left| \mathbf{\nabla }\Phi ^{t+\varepsilon }\right| }{2}\mathbf{V}%
^{\varepsilon }\right\| _{L^{2}}^{2}\leq \left\| \tilde{q}_{\varepsilon
}\right\| _{L^{2}}\left\| \mathbf{V}^{\varepsilon }\right\| _{L^{2}}
\end{equation}%
It follows that $\mathbf{V}^{\varepsilon }$ is uniformly bounded with
respect to $\varepsilon $ in $B_{\Phi ^{t+\varepsilon }}^{1}$ for $%
\varepsilon $ small enough.

Next, observe that 
\begin{equation}
\left( \mathbf{-\Delta +}\frac{\left| \mathbf{\nabla }\Phi ^{t}\right| ^{2}}{%
4}\right) \mathbf{V}^{\varepsilon }=\hat{q}_{\varepsilon }
\end{equation}%
where 
\begin{equation}
\hat{q}_{\varepsilon }=\tilde{q}_{\varepsilon }-\frac{\left| \mathbf{\nabla }%
\Phi ^{t+\varepsilon }-\mathbf{\nabla }\Phi ^{t}\right| ^{2}}{4}\mathbf{V}%
^{\varepsilon }+\frac{\left( \mathbf{\nabla }\Phi ^{t+\varepsilon }-\mathbf{%
\nabla }\Phi ^{t}\right) \cdot \mathbf{\nabla }\Phi ^{t}}{2}\mathbf{V}%
^{\varepsilon }
\end{equation}%
is uniformly bounded in $B^{0}$ with respect to $\varepsilon $ for small
enough $\varepsilon .$ Using \ regularity, it follows that for small enough $%
\varepsilon ,$ $\mathbf{V}^{\varepsilon }$ is uniformly bounded in $B_{\Phi
^{t}}^{2}$ with respect to $\varepsilon .$This implies that $\hat{q}%
_{\varepsilon }$ is uniformly bounded in $B_{\Phi ^{t}}^{1}$ for $%
\varepsilon $ small enough. Again, we can continue by a bootstrap argument
to consequently get that for $\varepsilon $ small enough, $\mathbf{V}%
^{\varepsilon }$ is uniformly bounded in $B_{\Phi ^{t}}^{k}$ for any $k.$%
\newline
It is then clear that \ for small enough $\varepsilon ,$ the right hand
sides of $\left( 23\right) $ is $\mathcal{O}(\varepsilon )$ in $B^{0}$ and
consequently, using the same argument as above, we get that $\left( \mathbf{w%
}(t)-\mathbf{v}^{\varepsilon }(t)\right) e^{-\Phi ^{t}/2}$ is $\mathcal{O}%
(\varepsilon )$ in $B_{\Phi ^{t}}^{2}$; again iterating the regularity
argument, we obtain that for small enough $\varepsilon ,$ $\left( \mathbf{w}%
(t)-\mathbf{v}^{\varepsilon }(t)\right) e^{-\Phi ^{t}/2}$ is $\mathcal{O}%
(\varepsilon )$ $B_{\Phi ^{t}}^{k}$ for every $k.$ We have proved:

\begin{proposition}
Under the above on $\Phi $ and $g,$ there exists $T>0$ so that for each $%
t\in (0,T),$ $\mathbf{v}^{\varepsilon }(t)$ converges to $\mathbf{w}(t)$ in $%
C^{\infty }.$
\end{proposition}

\begin{remark}
The proposition establishes that $\mathbf{v}(t)$ is differentiable in $t$
and $\dfrac{d}{dt}\mathbf{v}(t)$ is given by the unique $C^{\infty }-$%
solution $\mathbf{w}(t)$ of the system 
\begin{equation}
-\mathbf{\Delta w}(t)+\mathbf{\nabla }\Phi ^{t}\cdot \mathbf{\nabla w}(t)+%
\mathbf{Hess}\Phi ^{t}\mathbf{w}(t)=\mathbf{Hess}g\mathbf{v}(t)-\mathbf{%
\nabla }g\cdot \mathbf{\nabla v}(t).
\end{equation}%
Iterating this argument, we easily get that, $\mathbf{v}(t)$ is smooth in $%
t\in (0,T).$
\end{remark}

Now we are ready for the following:

\section{Formula for $\protect\theta ^{(n)}(t)$}

For an arbitrary suitable function $\ f(t)=f(t,w)$%
\begin{equation}
\frac{\partial }{\partial t}<f(t)>_{t,\Lambda }=<f^{\;\prime
}(t)>_{t,\Lambda }+\mathbf{cov}(f,g).
\end{equation}%
Hence,%
\begin{equation}
\frac{\partial }{\partial t}<f(t)>_{t,\Lambda }=<f^{\;\prime
}(t)>_{t,\Lambda }+<A_{\Phi ^{t}}^{(1)^{-1}}\left( \mathbf{\nabla }f\right)
\cdot \mathbf{\nabla }g>_{t,\Lambda }.
\end{equation}%
Let 
\begin{equation}
A_{g}f:=A_{\Phi ^{t}}^{(1)^{-1}}\left( \mathbf{\nabla }f\right) \cdot 
\mathbf{\nabla }g.
\end{equation}%
Thus,%
\begin{equation}
\frac{\partial }{\partial t}<f(t)>_{t,\Lambda }=<\left( \dfrac{\partial }{%
\partial t}+A_{g}\right) f>_{t,\Lambda }.
\end{equation}%
The linear operator $\dfrac{\partial }{\partial t}+A_{g}$ will be denoted by 
$H_{g}.$%
\begin{eqnarray*}
\theta _{\Lambda }^{\prime }(t) &=&<g>_{t,\Lambda } \\
&=&<\left( \dfrac{\partial }{\partial t}+A_{g}\right) ^{0}g>_{t,\Lambda } \\
&=&<H_{g}^{0}g>_{t,\Lambda };
\end{eqnarray*}%
\begin{eqnarray*}
\theta _{\Lambda }^{\prime \prime }(t) &=&\frac{\partial }{\partial t}%
<g>_{t,\Lambda } \\
&=&<A_{\Phi ^{t}}^{(1)^{-1}}\left( \mathbf{\nabla }g\right) \cdot \mathbf{%
\nabla }g>_{t,\Lambda } \\
&=&<\left( \dfrac{\partial }{\partial t}+A_{g}\right) g>_{t,\Lambda };
\end{eqnarray*}%
\begin{eqnarray*}
\theta _{\Lambda }^{\prime \prime \prime }(t) &=&\frac{\partial }{\partial t}%
<A_{\Phi ^{t}}^{(1)^{-1}}\left( \mathbf{\nabla }g\right) \cdot \mathbf{%
\nabla }g>_{t,\Lambda } \\
&=&<\frac{\partial }{\partial t}\left( A_{\Phi ^{t}}^{(1)^{-1}}\left( 
\mathbf{\nabla }g\right) \cdot \mathbf{\nabla }g\right) >_{t,\Lambda } \\
+ &<&\left( A_{\Phi ^{t}}^{(1)^{-1}}\mathbf{\nabla }\left( A_{\Phi
^{t}}^{(1)^{-1}}\left( \mathbf{\nabla }g\right) \cdot \mathbf{\nabla }%
g\right) \right) \cdot \mathbf{\nabla }g>_{t,\Lambda } \\
&=&<\left( \dfrac{\partial }{\partial t}+A_{g}\right) ^{2}g>_{t,\Lambda }.
\end{eqnarray*}%
By induction it is easy to see that%
\begin{eqnarray*}
\theta _{\Lambda }^{(n)}(t) &=&<\left( \dfrac{\partial }{\partial t}%
+A_{g}\right) ^{n-1}g>_{t,\Lambda }\;\;\;\;\;\;\;(\forall n\geq 1) \\
&=&<H_{g}^{(n-1)}g>_{t,\Lambda }.
\end{eqnarray*}%
Next, we propose to find a simpler formula for $\theta _{\Lambda }^{(n)}(t)$
that only involves $A_{g}.$%
\begin{eqnarray*}
H_{g}g &=&A_{\Phi ^{t}}^{(1)^{-1}}\left( \mathbf{\nabla }g\right) \cdot 
\mathbf{\nabla }g \\
&=&A_{g}g
\end{eqnarray*}%
\begin{equation}
H_{g}^{2}g=\frac{\partial }{\partial t}\mathbf{\nabla }f\cdot \mathbf{\nabla 
}g+\left( A_{\Phi ^{t}}^{(1)^{-1}}\mathbf{\nabla }\left( A_{\Phi
^{t}}^{(1)^{-1}}\left( \mathbf{\nabla }g\right) \cdot \mathbf{\nabla }%
g\right) \right) \cdot \mathbf{\nabla }g
\end{equation}%
where $f$ satisfies the equation 
\begin{equation}
\mathbf{\nabla }f=A_{\Phi ^{t}}^{(1)^{-1}}\left( \mathbf{\nabla }g\right) .
\end{equation}%
With $\mathbf{v}(t)=\mathbf{\nabla }f,$ as before, we get 
\begin{equation*}
\frac{\partial }{\partial t}\mathbf{\nabla }f\cdot \mathbf{\nabla }g=A_{\Phi
^{t}}^{(1)^{-1}}\left( \mathbf{Hess}g\mathbf{v}(t)-\mathbf{\nabla }g\cdot 
\mathbf{\nabla v}(t)\right) \cdot \mathbf{\nabla }g
\end{equation*}%
and $H_{g}^{2}$ becomes 
\begin{eqnarray*}
H_{g}^{2}g &=&A_{\Phi ^{t}}^{(1)^{-1}}\left[ \left( \mathbf{Hess}g\mathbf{v}%
(t)-\mathbf{\nabla }g\cdot \mathbf{\nabla v}(t)\right) +\mathbf{\nabla }%
\left( A_{\Phi ^{t}}^{(1)^{-1}}\left( \mathbf{\nabla }g\right) \cdot \mathbf{%
\nabla }g\right) \right] \cdot \mathbf{\nabla }g \\
&=&A_{\Phi ^{t}}^{(1)^{-1}}2\mathbf{\nabla }\left( A_{g}g\right) \cdot 
\mathbf{\nabla }g \\
&=&2A_{g}^{2}g.
\end{eqnarray*}

\begin{proposition}
\textit{If }%
\begin{equation*}
\theta _{\Lambda }(t)=\log \left[ \int dxe^{-\Phi ^{t}(x)}\right]
\end{equation*}%
\textit{where}%
\begin{equation*}
\Phi ^{t}(x)=\Phi _{\Lambda }(x)-tg(x)
\end{equation*}%
\textit{is as above then }$\theta _{\Lambda }^{(n)}(t),$\textit{\ the }$nth-$%
\textit{\ derivative of }$\theta _{\Lambda }(t)$\textit{\ is given by the
formula }%
\begin{equation*}
\theta _{\Lambda }^{\prime }(t)=<g>_{t,\Lambda },
\end{equation*}%
\textit{and for }$n\geq 1$%
\begin{equation*}
\theta _{\Lambda }^{(n)}(t)=\left( n-1\right) !<A_{g}^{n-1}g>_{t,\Lambda }.
\end{equation*}
\end{proposition}

\begin{proof}
We have already established that 
\begin{equation*}
\theta _{\Lambda }^{(n)}(t)=<H_{g}^{n-1}g>_{t,\Lambda }\;\;\;\;for\;n\geq 1.
\end{equation*}%
It then only remains to prove that 
\begin{equation*}
H_{g}^{n-1}g=\left( n-1\right) !A_{g}^{n-1}g\;\;\;\;\;\;for\;n\geq 1.
\end{equation*}%
The result is already established above for $n=1,2,3,.$ By induction, assume
that 
\begin{equation*}
H_{g}^{n-1}g=\left( n-1\right) !A_{g}^{n-1}g\;.
\end{equation*}%
if $n$ is replaced by $\tilde{n}\leq n.$%
\begin{eqnarray*}
H_{g}^{n}g &=&\left( \dfrac{\partial }{\partial t}+A_{g}\right) \left(
\left( n-1\right) !A_{g}^{n-1}g\right) \\
&=&\left( n-1\right) !\left( \dfrac{\partial }{\partial t}%
A_{g}^{n-1}g+A_{g}^{n}g\right) .
\end{eqnarray*}%
Now 
\begin{eqnarray*}
A_{g}^{n-1}g &=&\left[ A_{\Phi ^{t}}^{(1)^{-1}}\mathbf{\nabla }\left(
A_{g}^{n-2}g\right) \right] \cdot \mathbf{\nabla }g\; \\
&=&\mathbf{\nabla }\varphi _{n}\cdot \mathbf{\nabla }g
\end{eqnarray*}%
where%
\begin{equation*}
\mathbf{\nabla }\varphi _{n}=\left[ A_{\Phi ^{t}}^{(1)^{-1}}\mathbf{\nabla }%
\left( A_{g}^{n-2}g\right) \right] .
\end{equation*}%
We obtain,%
\begin{equation*}
\dfrac{\partial }{\partial t}\mathbf{\nabla }\varphi _{n}=A_{\Phi
^{t}}^{(1)^{-1}}\left( \dfrac{\partial }{\partial t}\mathbf{\nabla }%
A_{g}^{n-2}g+\mathbf{Hess}g\mathbf{\nabla }\varphi _{n}-\mathbf{\nabla }%
g\cdot \mathbf{\nabla }\left( \mathbf{\nabla }\varphi _{n}\right) \right) .
\end{equation*}%
We then have 
\begin{eqnarray*}
\dfrac{\partial }{\partial t}A_{g}^{n-1}g &=&\dfrac{\partial }{\partial t}%
\mathbf{\nabla }\varphi _{n}\cdot \mathbf{\nabla }g \\
&=&\left[ A_{\Phi ^{t}}^{(1)^{-1}}\left( \dfrac{\partial }{\partial t}%
\mathbf{\nabla }A_{g}^{n-2}g+\mathbf{Hess}g\mathbf{\nabla }\varphi _{n}-%
\mathbf{\nabla }g\cdot \mathbf{\nabla }\left( \mathbf{\nabla }\varphi
_{n}\right) \right) \right] \cdot \mathbf{\nabla }g \\
&=&\left[ A_{\Phi ^{t}}^{(1)^{-1}}\left( \dfrac{\partial }{\partial t}%
\mathbf{\nabla }A_{g}^{n-2}g+\mathbf{\nabla }\left( \mathbf{\nabla }\varphi
_{n}\cdot \mathbf{\nabla }g\right) \right) \right] \cdot \mathbf{\nabla }g \\
&=&A_{g}\left[ \dfrac{\partial }{\partial t}A_{g}^{n-2}g+A_{g}\left(
A_{g}^{n-2}g\right) \right] \\
&=&A_{g}H_{g}\left( A_{g}^{n-2}g\right) . \\
&=&A_{g}H_{g}\left( \frac{1}{\left( n-2\right) !}H_{g}^{(n-2)}g\right)
\;\;\;\;\;\;\;(\text{from the induction hypothesis}) \\
&=&\frac{1}{\left( n-2\right) !}A_{g}H_{g}^{(n-1)}g\; \\
&=&\frac{1}{\left( n-2\right) !}A_{g}\left( \left( n-1\right)
!A_{g}^{n-1}g\right) \;\;\;\;\;(\text{still by the induction hypothesis})\;\
\ \ \  \\
&=&(n-1)A_{g}^{n}g.
\end{eqnarray*}%
Thus,%
\begin{eqnarray*}
H_{g}^{n}g &=&\left( n-1\right) !\left( n-1+1\right) A_{g}^{n}g \\
&=&n!A_{g}^{n}g
\end{eqnarray*}
\end{proof}

\begin{proposition}
\textit{If }$g(0)=0,$ \textit{then }the formula 
\begin{equation*}
\theta _{\Lambda }^{(n)}(t)=\left( n-1\right) !<A_{g}^{n-1}g>_{t,\Lambda
},\;\;\;n\geq 2
\end{equation*}%
\newline
still holds if we no longer require $\Psi $ and $g$ to be compactly
supported in $\mathbb{R}^{\Lambda }.$
\end{proposition}

\begin{proof}
As in [8], consider the family cutoff functions 
\begin{equation}
\chi =\chi _{\varepsilon }
\end{equation}%
$(\varepsilon \in \lbrack 0,1])$ in $\mathcal{C}_{o}^{\infty }(\mathbb{R})$
with value in $[0,1]$ such that%
\begin{equation*}
\left\{ 
\begin{array}{c}
\chi =1\text{ \ \ \ \ \ \ \ \ \ \ \ \ \ \ \ \ for }\left| t\right| \leq
\varepsilon ^{-1}\text{ } \\ 
\left| \chi ^{(k)}(t)\right| \leq C_{k}\dfrac{\varepsilon }{\left| t\right|
^{k}}\text{ \ \ \ \ \ \ \ \ \ \ \ \ \ \ \ \ for }k\in \mathbb{N\ }\text{\ \
\ \ \ \ \ \ \ \ \ \ \ \ \ \ \ \ \ \ \ \ \ }%
\end{array}%
\right.
\end{equation*}%
We could take for instance%
\begin{equation*}
\chi _{\varepsilon }(t)=f(\varepsilon \ln \left| t\right| )
\end{equation*}%
for a suitable $f$.

We then introduce%
\begin{equation}
\Psi _{\varepsilon }(x)=\chi _{\varepsilon }(\left| x\right| )\Psi ,\ \ \ \
\ \ \ \ \ x\in \mathbb{R}^{\Lambda }
\end{equation}%
and 
\begin{equation}
g_{\varepsilon }(x)=\chi _{\varepsilon }(\left| x\right| )g\text{ \ \ \ \ \
\ \ \ \ \ \ \ }x\in \mathbb{R}^{\Gamma }\text{\ }
\end{equation}%
One can check that both $\Psi _{\varepsilon }(x)$ and $g_{\varepsilon }(x)$
satisfies the assumptions made above on $\Psi $ and $g.$ Now consider the
equation 
\begin{equation}
-\mathbf{\Delta }f_{\varepsilon }+\mathbf{\nabla }\Phi _{\varepsilon
}^{t}\cdot \mathbf{\nabla }f_{\varepsilon }=g_{\varepsilon }-<g_{\varepsilon
}>_{t,\Lambda .}
\end{equation}%
which implies 
\begin{equation}
\left( -\mathbf{\Delta }+\mathbf{\nabla }\Phi _{\varepsilon }^{t}\cdot 
\mathbf{\nabla }\right) \otimes \mathbf{v}_{\varepsilon }+\mathbf{Hess}\Phi
_{\varepsilon }^{t}\mathbf{v}_{\varepsilon }=\mathbf{\nabla }g_{\varepsilon }
\end{equation}%
where 
\begin{equation*}
\mathbf{v}_{\varepsilon }\mathbf{=\nabla }f_{\varepsilon }
\end{equation*}%
It was proved in [8] that $\mathbf{v}_{\varepsilon }=A_{\Phi ^{t}}^{(1)^{-1}}%
\mathbf{\nabla }g_{\varepsilon }\;$converges in $C^{\infty }$ to $A_{\Phi
^{t}}^{(1)^{-1}}\mathbf{\nabla }g$ as $\varepsilon \rightarrow 0.$
\end{proof}

\begin{remark}
If we denote by $R_{n}$ the remainder of the Taylor series expansion of the
pressure $P_{\Lambda }(t),$ given by 
\begin{equation*}
P_{\Lambda }(t)=\dfrac{\theta _{\Lambda }(t)}{\left| \Lambda \right| }
\end{equation*}%
we have 
\begin{eqnarray*}
R_{n} &=&\frac{P_{\Lambda }^{(n+1)}(t_{o})}{(n+1)!} \\
&=&\left. \dfrac{<A_{g}^{n}g>_{t,\Lambda }}{\left( n+1\right) \left| \Lambda
\right| }\right| _{t=t_{o}}.
\end{eqnarray*}%
If $\Phi $ and $g$ are such that $<A_{g}^{n}g>_{t,\Lambda }$ is uniformly
bounded with respect to $n$ and does not grow faster than $\left| \Lambda
\right| $, we automatically get the analyticity of the pressure in the
thermodynamic limit.
\end{remark}

\section{Some Consequences of the Formula for $nth-$Derivative of the
Pressure.}

In the following, we shall additionally assume that 
\begin{equation*}
\left. 
\begin{array}{c}
\mathbf{\nabla }g(0)=0,\;\;\;\;\;\text{and} \\ 
\mathbf{\nabla }\Phi _{\Lambda }^{t}(0)=0\text{ \ \ \ for all }t\in \lbrack
0,T).%
\end{array}%
\right.
\end{equation*}%
When $n=1,$ we recall that $A_{g}^{0}g=g$, 
\begin{equation*}
\theta _{\Lambda }^{\prime }(t)=<g>_{t,\Lambda }
\end{equation*}%
and if 
\begin{equation*}
\mathbf{v}(t)=\mathbf{\nabla }f=A_{\Phi ^{t}}^{(1)^{-1}}\mathbf{\nabla }g,
\end{equation*}%
then we have 
\begin{equation*}
\left( -\mathbf{\Delta }+\mathbf{\nabla }\Phi _{\Lambda }^{t}\cdot \mathbf{%
\nabla }\right) \otimes \mathbf{v}(t)+\mathbf{Hess}\Phi _{\Lambda }^{t}%
\mathbf{v}(t)=\mathbf{\nabla }g
\end{equation*}%
and as in [8] $\mathbf{v}(t)$ is a solution of the equation%
\begin{equation}
g=<g>_{t,\Lambda }+\mathbf{v}(t)\cdot \mathbf{\nabla }\Phi _{\Lambda
}^{t}-div\mathbf{v}(t).
\end{equation}%
Using the assumptions above, we have 
\begin{eqnarray*}
\theta _{\Lambda }^{\prime }(t) &=&<g>_{t,\Lambda } \\
&=&div\mathbf{v}(t)(0).
\end{eqnarray*}%
Similarly, the formula 
\begin{equation*}
\theta _{\Lambda }^{(n)}(t)=\left( n-1\right) !<A_{g}^{n-1}g>_{t,\Lambda },
\end{equation*}%
implies that 
\begin{equation*}
\theta _{\Lambda }^{(n)}(t)=\left( n-1\right) !div\mathbf{v}_{n}(t)(0),
\end{equation*}%
where 
\begin{equation*}
\mathbf{v}_{n}(t)=A_{\Phi ^{t}}^{(1)^{-1}}\mathbf{\nabla }\left(
A_{g}^{n-1}g\right) .
\end{equation*}

\begin{remark}
The idea of representing $\theta _{\Lambda }^{\prime }(t)$ in terms of $div%
\mathbf{v}(t)(0)$ is due to Helffer and Sj\"{o}strand [8] in the context of
proving the exponential convergence of the thermodynamic limit in the one
dimensional case.
\end{remark}

We conclude these notes by a discussion about the potential contribution of
this results towards solving the two dimensional dipole gas problem. The
dipole gas and other gases of particles interacting through Coulomb forces
are very important statistical systems. In particular, for dipole gas, the
lack of screening is well known [59], and the analyticity of the pressure in
the high temperature and low activity region has been proved in an indirect
way, by means of renormalization group methods (see [60] and [61]).\newline
A direct proof of the analyticity of the pressure based on estimating the
coefficients of the Mayer (Taylor) series is still an open problem. The
close relationship between this model and the Coulomb gas in the
Kostelitz-Thouless phase ($\beta >8\pi $), go along with the non-existence
of any proof for the analyticity of the pressure in the Coulomb gas.
Indirect arguments are attempted in [62],[63] and [64]. We believe that
after a suitable \ regularization of the Coulomb potential at short
distances to assure stability, we can fit the problem into the framework of
the model discussed above and get an estimate of the coefficients of the
Mayer series through our formula for the n$th$ derivative of the pressure.

\textbf{Acknowledgements:} I would like to thank my advisor Haru Pinson for
all the fruitful discussions and the help he has provided in the writing of
these notes. I also would like to thank Prof. Tom Kennedy, Prof. William
Faris, and all members of the mathematical physics group at the University
of Arizona. Special thanks also goes to Prof. Kenneth \ D. McLaughlin for
accepting to discuss with me about the ideas developed in this paper.


\begin{thebibliography}{99}
\bibitem{} Brascamp. H and J and Lieb. E. H, \textit{On extensions of the
Brunn-Minkowski and Prekopa-Leindler theorems including inequalities for log
concave\ functions, and with application to the diffusion equation, J.
Funct. Analysis, 22 (1976), 366-389.}

\bibitem{} Bodineau. T and Helffer. B, \textit{Correlations, spectral gap
and logSobolev inequalities for unbounded spins sytems, Proc. UAB Conf.
March 16-20\ 1999, AMS/IP stud. adv. math 16 (2000), 51-66.}

\bibitem{} Evans. L. C, \textit{Partial Differential Equations''\ (AMS,
1998).}

\bibitem{} Helffer. B, \textit{Introduction to the semiclassical analysis
for the schrodinger operator \ and applications, Lecture Notes in Math, 1336
(1988).}

\bibitem{} Helffer. B, \textit{Around a stationary phase theorem in large
dimension. J. Funct. Anal. 119 (1994), no. 1, 217-252.}

\bibitem{} Helffer. B, \textit{Semiclassical analysis, Witten laplacians and
statistical mechanics series on partial differential equations and
applications-Vol.1 - World Scientific (2002).}

\bibitem{} Helffer. B, \textit{Remarks on decay of correlations and Witten
laplacians. II, analysis of the dependence on the interaction. Rev. Math.
Phys, 11 (1999), no.\ 3, 321-336}

\bibitem{} Helffer. B and Sj\"{o}strand. J, \textit{On the correlation for
Kac-like models in the convex case. J. of Stat. phys, 74 Nos.1/2, 1994.}

\bibitem{} Helffer. B and Sj\"{o}strand. J, \textit{Semiclassical expansions
of the thermodynamic limit for a Schr\"{o}dinger equation. The one well
case. M\'{e}thodes semi-classiques, Vol. 2 (Nantes, 1991). Ast\'{e}risque
No. 210 (1992), 7-8, 135-181.}

\bibitem{} Johnsen, Jon: \textit{On the spectral properties of
Witten-Laplacians, their range projections and Brascamp-Lieb's inequality.
Integral Equations Operator Theory 36 (2000),\ no. 3, 288-324.}

\bibitem{} Kneib and Jean-Marie Mignot, \textit{Fulbert \'{E}quation de
Schmoluchowski g\'{e}n\'{e}ralis\'{e}e. (French) [generalized Smoluchowski
equation] Ann. Mat. Pura\ Appl. (4) 167 (1994), 257-298.}

\bibitem{} Naddaf. A and Spencer. T, \textit{On homogenization and scaling
limit of gradient perturbations of a massless free field, Comm. Math.
Physics\ 183 (1997), 55-84.}

\bibitem{} Sj\"{o}strand. J, \textit{Correlation asymptotics and Witten
laplacians, Algebra and Analysis 8, no. 1 (1996), 160-191.}

\bibitem{} Sj\"{o}strand. J, \textit{Exponential convergence of the first
eigenvalue divided by the dimension, for certain sequences of Schr\"{o}%
dinger operators. M\'{e}thodes semi-classiques, Vol. 2 (Nantes, 1991). Ast%
\'{e}risque No. 210 (1992), 10, 303-326.}

\bibitem{} Sj\"{o}strand, J, \textit{Potential wells in high dimensions. II.
More about the one well case. Ann. Inst. H. Poincar\'{e} Phys. Th\'{e}or.
58, no. 1 (1993), 43-53.}

\bibitem{} Sj\"{o}strand. J, \textit{Potential wells in high dimensions. I.
Ann. Inst. H. Poincar\'{e} Phys. Th\'{e}or. 58, no. 1 (1993), 1-41.}

\bibitem{} Yosida. K, \textit{Functional analysis, springer classics in
mathematics by Kosaku Yosida.}

\bibitem{} Witten. E, \textit{Supersymmetry and Morse theory, J. of \ Diff.
Geom. 17, (1982), 661-692.}

\bibitem{} Cartier. P, \textit{Inegalit\'{e}s de corr\'{e}lation en m\'{e}%
canique statistique, S\'{e}minaire Bourbaki 25\'{e}me ann\'{e}e, 1972-1973,
No 431.}

\bibitem{} Kac. M, \textit{Mathematical mechanism of phase
transitions(Gordon and Breach, New York, 1966).}

\bibitem{} Troianiello. G. M, \textit{Elliptic Differential Equations and
Obstacle Problems (Plenum Press, New York 1987).}

\bibitem{} Berezin. F. A and Shubin. M. A, \textit{The Schr\"{o}dinger
Equation (Kluwer Academic Publisher, 1991).}

\bibitem{} Dobrushin. R. L, \textit{The description of random field by means
of conditional probabilities and conditions of its regularity.
Theor.Prob.Appl. 13, (1968), 197-224.}

\bibitem{} Dobrushin. R. L, \textit{Gibbsian random fields for lattice
systems with pairwise interactions. Funct. Anal. Appl. 2 (1968), 292-301.}

\bibitem{} Dobrushin. R. L, \textit{The problem of uniqueness of a Gibbs
random field and the problem of phase transition. Funct. Anal. Appl. 2
(1968), 302-312.}

\bibitem{} Bach. V, Jecko. T and Sjostrand. J, \textit{Correlation
asymptotics of classical lattice spin systems with nonconvex Hamilton
function at low temperature. Ann. Henri Poincare (2000), 59-100.}

\bibitem{} Bach. V and Moller. J. S, \textit{Correlation at low temperature,
exponential decay. Jour. funct. anal 203 (2003), 93-148.}

\bibitem{} Yang. C. N and Lee. T.D, \textit{\ Statistical theory of
equations of state and phase transition I. Theory of condensation. Phys.Rev.
87 (1952), 404-409.}

\bibitem{} Heilmann. O. J, \textit{Zeros of the grand partition function for
a lattice gas. J.Math.Phys. 11 (1970), 2701-2703.}

\bibitem{} Asano. T, \textit{Theorem on the partition functions of the
Heisenberg ferromagnets. J. Phys. Soc. Jap. 29 (1970), 350-359.}

\bibitem{} Ruelle. D, \textit{An Extension of lee-Yang circle theorem. Phys.
Rev. Letters, 26 (1971), 303-304.}

\bibitem{} Ruelle. D, \textit{Some remarks on the location of zeroes of the
partition function for lattice systems. Commun. Math. Phys 31, (1973),
265-277.}

\bibitem{} Slawny. J, \textit{Analyticity and uniqueness for spin 1/2
classical ferromagnetic lattice systems at low temperature Commun. Math.
Phys. 34 (1973), 271-296.}

\bibitem{} Gruber. C, Hintermann. A, and Merlini. D, \textit{Analyticity and
uniqueness of the invariant equilibrium state for general spin 1/2 classical
lattice spin systems. Commun. Math. Phys. 40 (1975), 83-95.}

\bibitem{} Griffiths. R. B, \textit{Rigorous results for Ising ferromagnets
of arbitrary spin. J. Math. Phys. 10 (1969), 1559-1565.}

\bibitem{} Simon. B and Griffiths. R. B, \textit{The }$\left( \Phi
^{4}\right) _{2}$\textit{\ Field theory as a classical Ising model. Commun.
Math. Phys. 33, (1973), 145-164.}

\bibitem{} Newman. C. M, \textit{Zeros of the partition function for
generalized Ising systems. Commun. Pure. Appl. Math. 27, (1974), 143-159.}

\bibitem{} Dunlop. F, \textit{Zeros of the partition function and gaussian
inequalities for the plane rotator model. J. Stat. Phys. 21 (1979), 561-572.}

\bibitem{} Dunlop. F, \textit{Analyticity of the pressure for Heisenberg and
plane rotor models. Commun. Math. Phys. 69 (1979), 81-88.}

\bibitem{} Lieb. E and Sokal. A. D, \textit{A general Lee-Yang theorem for
one-component and multicomponent ferromagnets. Commun. Math. Phys. 80
(1981), 153-179.}

\bibitem{} Glimm. J and Jaffe. A, \textit{Quantum Physics. A functional
integral point of view. New York ect. Springer (1981)}

\bibitem{} Bricmont. J, Lebowitz. J. L and Pfister. C. E, \textit{Low
temperature expansion for continuous spin Ising models. Commun. Math. Phys.
78 (1980), 117-135.}

\bibitem{} Dobrushin. R. L, \textit{Induction on volume and no Cluster
expansion. In: M. Mebkhout and R. Seneor (eds), VIII. Internat. Congress on
Mathematical Physics, Marseille 1986, Singapore: World Scientific, pp. 73-91.%
}

\bibitem{} Dobrushin. R. L and Sholsmann. S. B, \textit{Completely
analytical Gibbs fields. In: J. Fritz, A.Jaffe, and D.Sz\'{a}sz (eds)
Statistical Mechanics and Dynamical Systems, Boston ect. Birkh\"{a}user,
(1985), pp. 371-403.}

\bibitem{} Dobrushin. R.L and Sholsmann. S. B, \textit{Completely analytical
interactions: constructive description. J. Stat. Phys. 46 (1987), 983-1014.}

\bibitem{} Duneau. M, Iagolnitzer. D and Souillard. B, \textit{Decrease
properties of truncated correlation functions and analyticity properties for
classical lattice and continuous systems. Commun. Math. phys. 31 (1973),
191-208.}

\bibitem{} Duneau. M and Iagolnitzer. D and Souillard. B, \textit{Strong
cluster properties for classical systems with finite range interaction \
Commun. Math. Phys. 35 (1974), 307-320.}

\bibitem{} Duneau. M and Iagolnitzer. D and Souillard. B, \textit{Decay of
correlations for infinite range interactions. J. Math. Phys. 16 (1975),
1662-1666.}

\bibitem{} Glimm. J and Jaffe. A, \textit{Expansion in Statistical Physics.
Commun. Pure. Appl. Math. 38 (1985), 613-630.}

\bibitem{} Israel. R. B, \textit{High temperature analyticity in classical
lattice systems. Commun. Math. Phys. 50 (1976), 245-257.}

\bibitem{} Koteck\'{y}. R and Preiss. D, \textit{Cluster expansions for
abstract polymers models. Commun. Math. Phys. 103, (1986), 491-498.}

\bibitem{} Kunz. H, \textit{Analyticity and clustering proporties of
unbounded spin systems. Commun. Math. Phys. 59 (1978), 53-69.}

\bibitem{} Lebowitz. J. L, \textit{Bounds on the correlations and
analyticity properties of Ising spin systems. Commun. Math. Phys. 28 (1972),
313-321.}

\bibitem{} Lebowit., J. L, \textit{\ Uniqueness, analyticity and decay
properties of correlations in equilibrium systems. In: H. Araki (ed)
International Symposium on Mathematical Problems in Theoretical Physiscs.
LNPH. 80 (1975), pp. 68-80.}

\bibitem{} Malyshev. V. A, \textit{Cluster expansions in lattice models of
statistical physics and the quantum theory of fields. Russian Math Surveys.
35,2 (1980), 3-53.}

\bibitem{} Malyshev. V. A and Milnos. R. A, \textit{Gibbs Random Fields: The
method of cluster expansions (In Russian) Moscow: Nauka (1985).}

\bibitem{} Prakash. C, \textit{High temperature differentiability of lattice
Gibbs states by Dobrushin uniqueness techniques. J. Stat.Phys, 31 (1983),
169-228.}

\bibitem{} Jost. J\"{u}rgen, \textit{Riemannian Geometry and Geometric
Analysis. 4th ed Berlin : Springer, c2005.}

\bibitem{} Park. Y. M, \textit{Lack of screening in the continuous dipole
systems, Comm. Math. Phys. 70 (1979), 161-167.}

\bibitem{} Gawedzki. K and Kupiainen. A, \textit{Block spin renormalization
group for dipole gas and }$\left( \nabla \phi \right) ^{4},$\textit{\ Ann.
Phys, (1983), 147-198.}

\bibitem{} Brydges. D and Yau. H. T, \textit{Grad }$\phi $\textit{\
perturbations of massless gaussian fields, Comm. Math. Phys. (1990), 129-351.%
}

\bibitem{} Fr\"{o}hlich. J and Spencer. T, \textit{On the statistical
mechanics of classical Coulomb and dipole gases, J. Stat. Phys. 24 (1981),
617-701.}

\bibitem{} \textit{\ }Fr\"{o}hlich. J and Park. Y. M, \textit{Correlation
inequalities in the thermodynamic limit for classical and quantum systems.
Comm. Math. Phys, 59 (1990), 235-266.}

\bibitem{} Marchetti. D. H and Klein. A, \textit{Power law fall-off in the
two dimensional Coulomb gases at inverse temperature }$\beta >8\pi ,$\textit{%
\ J.Stat.Phys. 64 (1991), 135.}

\bibitem{} Berezin. F. A and Shubin. M. A, \textit{The Schr\"{o}dinger
Equation (Kluwer Academic Publisher, 1991).}

\bibitem{} Lo. Assane, \textit{Witten laplacian methods for the decay of
correlations. Preprint (2006).}
\end{thebibliography}
\end{document}